\documentclass[3p,12pt]{elsarticle}
\usepackage{graphicx}
\usepackage{amssymb}

\begin{document}

\begin{frontmatter}

\title{Notes on van der Meer Scan for Absolute Luminosity Measurement}
\author[cern,itep]{Vladislav Balagura}
\address[cern]{CERN, CH-1211 Geneve 23, Switzerland}
\address[itep]{ITEP, 117218, B.Cheremushkinskaya 25, Moscow, Russia}
\ead{balagura@cern.ch}

\begin{abstract}
  An absolute luminosity can be measured in an accelerator by sweeping
  beams transversely across each other in the so called van der Meer
  scan.  We prove that the method can be applied in the general case
  of arbitrary beam directions and a separation scan plane.  A simple
  method to develop an image of the beam in its transverse plane from
  spatial distributions of interaction vertexes is also
  proposed. From the beam images one can determine their overlap and
  the absolute luminosity. This provides an alternative way of the
  luminosity measurement during van der Meer scan.
\end{abstract}

\begin{keyword}
luminosity \sep van der Meer scan \sep beam imaging
\end{keyword}

\end{frontmatter}

%\linenumbers

\section{Van der Meer method for arbitrary beam velocities} 
\label{sec:lumi}

The luminosity of an accelerator is given by
\begin{equation}
\label{eq:lumi}
L = f N_1 N_2 K \cdot\int\! \rho^{\mathrm{lab}}_1(\vec{r}-\Delta\vec{r},\,t)\rho^{\mathrm{lab}}_2(\vec{r},\,t)\, d^3\vec{r}\,dt,
\end{equation}
where $K=\sqrt{(\vec{v}_1 - \vec{v}_2)^2 -
  \frac{(\vec{v}_1\times\vec{v}_2)^2}{c^2}}$ is the M\o ller kinematic
relativistic factor~\cite{moller}, $c$ is the speed of light,
$N_{1,2}$ are the number of particles in the colliding bunches all
moving with the common velocities $\vec{v}_{1,2}$, $f$ is the
frequency of collisions and $\rho^{\mathrm{lab}}_{1,2}(\vec{r},t)$ are
the normalized particle densities in the laboratory frame, so that
$\int\!\rho^{\mathrm{lab}}_{1,2}(\vec{r},t)\,d^3\vec{r}=1$ at any time
$t$.  The absolute value of the luminosity or the cross section can be
measured by separating the beams in the transverse plane by
$\Delta\vec{r}$ and by monitoring a collision rate as a function of
$\Delta\vec{r}$. This method was proposed by van der Meer more than 40
years ago and was originally proved in~\cite{vdm} for arbitrary beam
shapes and parallel beams $\vec{v}_1 \| \vec{v}_2$.  It was
successfully applied with various modifications at ISR~\cite{vdm,isr},
RHIC~\cite{rhic} and recently at LHC~\cite{lhc} accelerators.  It was
often used in the approximation of Gaussian or double Gaussian beam
shapes. At RHIC, for example, this allowed to take into account
various corrections due to the so-called hourglass effect, the
beam-beam deflection and the beam crossing angle.  The latter alone,
however, does not require any significant changes in the original van
der Meer method. Since we did not see any publication on this subject,
in this section we present a proof of van der Meer formula in case of
arbitrary beam crossing angle and beam shapes. It is applicable to the
scans at LHC where hourglass and beam-beam effects are
small~\cite{lhc}.

\begin{figure}[htbp]
\begin{center}
\includegraphics[width=0.7\textwidth,angle=0]{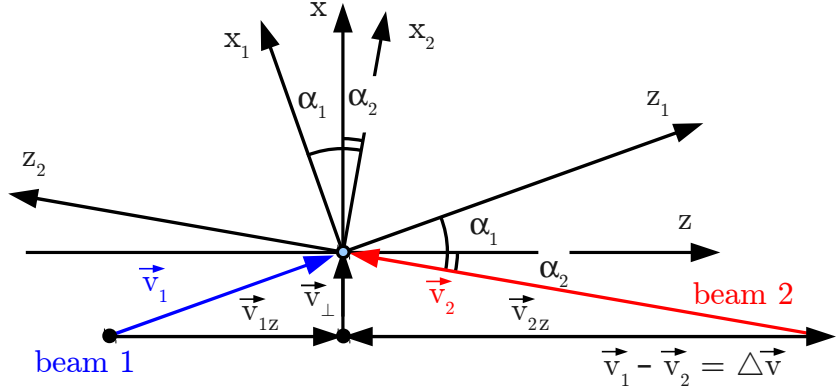}
\caption{Laboratory ($x$,
  $z$), first and second beam ($x_{1,2}$,
  $z_{1,2}$) coordinate systems in the crossing plane.}
\end{center}
\label{fig:coord}
\end{figure}

Without loss of generality in Eq.~(\ref{eq:lumi}) it is assumed that
only the first beam is moved.  We choose a coordinate system as shown
in Fig.~\ref{fig:coord} with $z$ axis along the direction
$\Delta\vec{v} = \vec{v}_1-\vec{v}_2$, $x$ axis lying in the beam
crossing plane and $y$ axis perpendicular to $x$ and $z$. Let's denote
$z$- and $x$-components of the velocities as $\vec{v}_{1,2z}$ and
$\vec{v}_\perp$, respectively, so that $\vec{v}_{1,2} = \vec{v}_\perp
+ \vec{v}_{1,2z}$.  The beam displacement plane is not necessarily
perpendicular to $z$, and in general $\Delta\vec{r}$ has three
components $(\Delta x,\Delta y,\Delta z)$. Its projection to $x$-$y$
plane will be denoted by $\Delta\vec{r}_\perp$.  For the particles
uniformly moving with the velocities $\vec{v}_{1,2}$ the time
evolution of their densities obeys the rule
$\rho^{\mathrm{lab}}_{1,2}(\vec{r},\,t) =
\rho^{\mathrm{lab}}_{1,2}(\vec{r}-\vec{v}_{1,2}t,\,0)$, therefore
\[\frac{L(\Delta\vec{r})}{f N_1 N_2 K} =
\int\! \rho^{\mathrm{lab}}_1(\vec{r}-\Delta\vec{r}-\vec{v}_\perp t -
\vec{v}_{1z}t,\,0)\cdot\rho^{\mathrm{lab}}_2(\vec{r}-\vec{v}_\perp t -
\vec{v}_{2z}t,\,0)\,\frac{\partial(\vec{r},\,t)}{\partial(\vec{r}-\vec{v}_\perp t,\,t)}
d^3(\vec{r}-\vec{v}_\perp t)\, dt=\]
\[= \int\! \left[\int\!
  \rho^{\mathrm{lab}}_1(\vec{r}_{\perp}-\Delta\vec{r}_{\perp},\, z',
  \, 0)\, dz'
  \cdot\int\!\rho^{\mathrm{lab}}_2(\vec{r}_{\perp},\,z'',\,0)\,dz''\right]
\frac{\partial(z,\,t)}{\partial(z',\,z'')}\,d^2\vec{r}_{\perp}=\]
\begin{equation}
\label{eq:split}
= \frac{1}{|\Delta\vec{v}|}
\int\! \rho^{\mathrm{lab},\,\perp}_1(\vec{r}_{\perp}-\Delta\vec{r}_{\perp})\rho^{\mathrm{lab},\,\perp}_2(\vec{r}_{\perp})\,d^2\vec{r}_{\perp},
\end{equation}
where we changed the integration variables to $\vec{r}-\vec{v}_\perp
t= (x-v_\perp t,y,z)=(\vec{r}_{\perp},\, z)$ and $z-\Delta z-
v_{1z}t=z'$, $z+ v_{2z}t=z''$, the corresponding Jacobians were
\[\frac{\partial(\vec{r},\,t)}{\partial(\vec{r}-\vec{v}_\perp t,\,t)}=1,
\quad \frac{\partial(z,\,t)}{\partial(z',\,z'')} =
1/|\Delta\vec{v}|.\]
We also used the notation
$\rho^{\mathrm{lab},\,\perp}_{1,2}(\vec{r}_{\perp})=\int\!\rho^{\mathrm{lab}}_{1,2}(\vec{r}_{\perp},\,z,\,0)\,dz$
for the particle density projections on the plane perpendicular to
$z\parallel \Delta\vec{v}$ at $t=0$.  For the case
$\vec{v}_1\|\vec{v}_2$ when $z\parallel\vec{v}_{1,2}$ we recover the
usual formula
\begin{equation}
  L(\Delta\vec{r}) = f N_1 N_2\int\! \rho^{\mathrm{lab},\,\perp}_1(\vec{r}_{\perp}-\Delta\vec{r}_{\perp})\rho^{\mathrm{lab},\,\perp}_2(\vec{r}_{\perp})\,d^2\vec{r}_{\perp}.
\label{eq:parallel}
\end{equation}

Following van der Meer method, Eq.~(\ref{eq:split}) should be
integrated over $\Delta\vec{r}$. In three-dimensional space an
equation of $\Delta\vec{r}$ plane can be written in the form
$\Delta\vec{r}\cdot\vec{n}=A$ where $A$ is some constant and
$\vec{n}=(\cos\theta_x,\,\cos\theta_y,\,\cos\theta_z)$ is the unit
normal, $\theta_{x,y,z}$ are the angles between $\vec{n}$ and the
axes.  Since Eq.~(\ref{eq:split}) does not depend on $\Delta z$, an
integration over $\Delta\vec{r}$ can be performed with the help of
$\delta$-function as
\begin{equation}
  \label{eq:delta}
  d^2 \Delta\vec{r} = \delta(\Delta\vec{r}\cdot\vec{n}-A)\,d\Delta x\,d\Delta y\,d\Delta z=
  \frac{d\Delta x\,d\Delta y}{\cos\theta_z} = \frac{d^2\Delta \vec{r}_\perp}{\cos\theta_z}.  
\end{equation}
In other words, the independence on $\Delta z$ allows to vary only
$\Delta\vec{r}_\perp$ projection of $\Delta\vec{r}$, and
$1/\cos\theta_z$ appears as a difference between the area on
$\Delta\vec{r}$ plane and its $x$-$y$ projection. Integration of the
density product in Eq.~(\ref{eq:split}) over $d\Delta x\, d\Delta y=
d^2 \Delta\vec{r}_{\perp}$ gives
\[\int\!\rho^{\mathrm{lab},\,\perp}_1(\vec{r}_{\perp}-\Delta\vec{r}_{\perp})\,d^2(\vec{r}_{\perp}-\Delta\vec{r}_{\perp})\times
\int\!\rho^{\mathrm{lab},\,\perp}_2(\vec{r}_{\perp}) \,d^2
\vec{r}_{\perp} = 1,\]
and we obtain
\begin{equation}
\label{eq:vdm}
\int\!\frac{L(\Delta\vec{r})}{f N_1 N_2}\,d^2\Delta\vec{r} = 
\frac{K}{\cos\theta_z|\Delta\vec{v}|} = \frac{1}{\cos\theta_z} \sqrt{1 - \frac{(\vec{v}_1\times\vec{v}_2)^2}{(\vec{v}_1 - \vec{v}_2)^2c^2}}\ .
\end{equation}

For the process with the cross section $\sigma$ (including a
reconstruction efficiency) the rate of events is given by
\begin{equation}
  \label{eq:rate}
R(\Delta \vec{r})=\sigma\cdot L(\Delta \vec{r}).
\end{equation}
If $f$, $N_{1,2}$ are measured during the scan, the monitoring of the
rate $R(\Delta \vec{r})$ allows to determine the cross section and
then the absolute luminosity
\begin{equation}
  \label{eq:vdmrates}
  \sigma = \cos\theta_z
  \left[1 - \frac{(\vec{v}_1\times\vec{v}_2)^2}{(\vec{v}_1 -
      \vec{v}_2)^2c^2}\right]^{-1/2} \cdot \int\!
  \frac{R(\Delta\vec{r})}{f N_1 N_2}\,d^2\Delta\vec{r} \ .
\end{equation}
This is the generalized van der Meer formula which is valid for any
crossing angle between the beams and for arbitrary displacement plane.
Note, that if the latter is not perpendicular to $z$,
$\cos\theta_z\ne1$, $\Delta z$ component of the displacement affects
the time of the interactions but not the luminosity.

The term with the velocities $\vec{v}_{1,2}$ in
Eq.~(\ref{eq:vdmrates}) is simply a $\gamma$-factor of the boost with
the velocity $\vec{v}_\perp$ (see Fig.~\ref{fig:coord}).  Indeed, it
can be calculated as follows
\[ v_\perp = v_1\sin\alpha_1 =
\frac{|\vec{v}_1\times(\vec{v}_1-\vec{v}_2)|}{|\vec{v}_1-\vec{v}_2|} =
\frac{|\vec{v}_1\times\vec{v}_2|}{|\vec{v}_1-\vec{v}_2|},\]
\begin{equation}
  \gamma_\perp = \frac{1}{\sqrt{1-(v_\perp/c)^2}}=\left[1 -
    \frac{(\vec{v}_1\times\vec{v}_2)^2}{(\vec{v}_1-\vec{v}_2)^2c^2}\right]^{-1/2}=
  \frac{|\Delta\vec{v}|}{K}.
\label{eq:gamma}
\end{equation}
Its appearance here can be understood from the following alternative
proof of Eq.~(\ref{eq:vdmrates}). It is based on a relativistic
invariance.  We start from Eq.~(\ref{eq:parallel}) used by van der
Meer in his original paper~\cite{vdm} and valid for the collinear
beams $\vec{v}_1\parallel\vec{v}_2$.  Integration over
$\Delta\vec{r}_\perp$ gives
\begin{equation}
  \sigma = \int\!\frac{(R/f)}{N_1 N_2}\,d^2\Delta\vec{r}_\perp^{\mathrm{\ coll}},
\label{eq:int_parallel}
\end{equation}
where ``coll'' superscript of $\Delta\vec{r}_\perp^{\mathrm{\ coll}}$
reminds us that the formula is valid only in the frame where the beams
are collinear.  The element $\Delta\vec{r}_\perp^{\mathrm{\ coll}}$ is
invariant only under boosts along $z$ which also preserve the
condition $\vec{v}_1\parallel\vec{v}_2$.  The ratio of two frequencies
$R/f$ is an average number of interactions per collision, which does
not depend on the choice of the coordinate system.  $\sigma$ is also a
relativistic invariant according to the cross section definition
proposed by M\o ller~\cite{moller}. The only non-invariant quantity in
Eq.~(\ref{eq:int_parallel}) is the displacement
$\Delta\vec{r}_\perp^{\mathrm{\ coll}}$.

The general case $\vec{v}_1\nparallel\vec{v}_2$ (see
Fig.~\ref{fig:coord}) can always be reduced to the collinear beams by
making a boost with the velocity $\vec{v}_\perp$. It is easy to show
that regardless of the laboratory $z$-components $\vec{v}_{1,2z}$, in
the relativistically boosted frame the beams become parallel to $z$,
so that Eq.~(\ref{eq:int_parallel}) is valid. To transform it back to
the laboratory frame we note that $d^2\Delta \vec{r}_\perp$ transforms
in the same way as a transverse part $dx\,dy$ of a relativistically
invariant four-dimensional volume $dx\,dy\,dz\,dt$. Since $dz$ is not
affected by the boost while $dt$ acquires a $\gamma$-factor, we have
$d^2\Delta \vec{r}_\perp = d^2\Delta \vec{r}_\perp^{\mathrm{\
    coll}}/\gamma_\perp$, so that the relativistically invariant
generalization of Eq.~(\ref{eq:int_parallel}) is
\begin{equation}
  \sigma = \gamma_\perp\int\!\frac{R(\Delta\vec{r})}{fN_1 N_2}\,d^2\Delta\vec{r}_\perp
  = \gamma_\perp\cos\theta_z\int\!\frac{R(\Delta\vec{r})}{fN_1 N_2}\,d^2\Delta\vec{r}.
\label{eq:int_any}
\end{equation}
Together with Eq.~(\ref{eq:gamma}) this completes the proof of
Eq.~(\ref{eq:vdmrates}).

%------------------------------------------------------------
For the following discussion it is useful to introduce coordinates
linked to the beams as shown in Fig.~\ref{fig:coord}. They are denoted
by ``$1,2$'' subscripts. $z_{1,2}$ axes are chosen along the beams,
$x_{1,2}$ axes lie in the beam crossing plane and other axes coincide,
$y_1=y_2=y$.  If $\alpha_{1,2}$ is the angle between $z_{1,2}$ and
$z$, we have
\begin{equation}
  \label{eq:frames}
  x = x_{1,2}\cos\alpha_{1,2}+ z_{1,2}\sin\alpha_{1,2},\quad z = \mp
  x_{1,2}\sin\alpha_{1,2}\pm z_{1,2}\cos\alpha_{1,2}\,.
\end{equation}

Let's assume that the distribution of particles in the transverse
$x$-$y$ plane is independent in $x$ and $y$.  In the beam's frame this
means an independence in $x_{1,2}\cos\alpha_{1,2}+
z_{1,2}\sin\alpha_{1,2}$ and $y$, which is usually ensured by an
absence of $x_{1,2}-y$ accelerator coupling and an independence of
longitudinal ($z_{1,2}$) and transverse ($y$) particle distributions.
With this assumption, the two-dimensional integral over
$\Delta\vec{r}_\perp$ can be reduced to the product of {\it
  one-dimensional integrals} along any $\Delta x=\Delta x_0$ and
$\Delta y=\Delta y_0$ lines. Indeed, if the transverse densities
factorize as $\rho^{\mathrm{lab},\,\perp}_{1,2}(x, y) =
\rho^{\mathrm{lab},\,\perp}_{1,2\ x}(x)\cdot
\rho^{\mathrm{lab},\,\perp}_{1,2\ y}(y)$, the luminosity and the rate
according to Eqs.~(\ref{eq:split}) and (\ref{eq:rate}) can also be
factorized into $x$- and $y$-dependent terms, $R(\Delta x, \Delta y) =
R_x(\Delta x)\cdot R_y(\Delta y)$, and its integral over
$d^2\Delta\vec{r}_\perp$ can be expressed as
\[\int\! R(\Delta x, \Delta y)\, d\Delta x\, d\Delta y =
\int\! R_x(\Delta x)\, d\Delta x \int\!  R_y(\Delta y)\, d\Delta y
\frac{R_x(\Delta x_0)R_y(\Delta y_0)}{R(\Delta x_0,\Delta y_0)} = \]
\begin{equation}
  = \frac{\int\! R(\Delta x, \Delta y_0)\, d\Delta x \cdot \int\! R(\Delta x_0, \Delta y)\, d\Delta y}{R(\Delta x_0,\Delta y_0)}.
\label{eq:fact}
\end{equation}

In Eq.~(\ref{eq:fact}) we considered the simple case when the
integration is performed over $\Delta x$-$\Delta y$ plane, so that
$\cos\theta_z=1$.  After substituting Eq.~(\ref{eq:fact}) into
(\ref{eq:int_any}), the cross section can be written as
\begin{equation}
\label{eq:res}
\sigma = \gamma_\perp\ 
\frac{\int\! R(\Delta x, \Delta y_0)\, d\Delta x \cdot \int\! R(\Delta x_0, \Delta y)\, d\Delta y}{fN_1N_2R(\Delta x_0,\Delta y_0)}.
\end{equation}
The integrals in the enumerator of Eq.~(\ref{eq:fact}) can be measured
in two one-dimensional scans over $\Delta x$ at fixed $\Delta y_0$ and
vice versa. There is no need to make a full scan in $\Delta x$-$\Delta
y$ plane. Note, that to get larger rates it is advantageous to keep
the beam separation $(\Delta x_0,\Delta y_0)$ small, but in general
Eq.~(\ref{eq:res}) is valid for arbitrary $(\Delta x_0,\Delta y_0)$.

In derivation of Eq.~\ref{eq:res} we do not use the fact that $x$ axis
lies in the crossing plane.  If the transverse distributions
$\rho^{\mathrm{lab},\,\perp}_{1,2}$ are independent in two {\it
  arbitrary} directions $x'$ and $y'$ in $\Delta\vec{r}_\perp$ plane,
the formula~(\ref{eq:res}) written in the primed coordinates still
remains valid, and $x'$ and $y'$ may be chosen as scan axes. If they
are not perpendicular but form an angle $\alpha_{x'y'}$, one should
also include the corresponding Jacobian $\sin\alpha_{x'y'}$. Finally,
the two scan axes may extend beyond the $x$-$y$ plane if their
projections still coincide with $x'$, $y'$ axes. Let's denote the
displacements along such scan axes as $\Delta x''$, $\Delta y''$ and
their inclination angles to $x$-$y$ plane as
$\alpha_{x'',y''}$. Taking into account that $d\Delta x'(y') =
\cos\alpha_{x''(y'')}d\Delta x''(y'')$, one obtains a more general
formula
\begin{equation}
\label{eq:genres}
\sigma =  \gamma_\perp\sin\alpha_{x'y'}\cos\alpha_{x''}\cos\alpha_{y''}\
\frac{\int\! R(\Delta x'', \Delta y''_0)\, d\Delta x'' \cdot \int\! R(\Delta x''_0, \Delta y'')\, d\Delta y''}{fN_1N_2R(\Delta x''_0,\Delta y''_0)}.
\end{equation}

\section{Reconstruction of individual beam profiles}

The density of interaction vertexes accumulated during time
$\Delta T$ is given by
\begin{equation}
  \frac{d^3 N_{\mathrm{vx}}}{dx\,dy\,dz}=f \Delta T N_1 N_2 K\sigma
  \int\!
  \rho^{\mathrm{lab}}_1(\vec{r}-\delta\vec{r}-\Delta\vec{r},\,t)\,\rho^{\mathrm{lab}}_2(\vec{r}-\delta\vec{r},\,t)\,
  V^{\mathrm{lab}}(\delta\vec{r})\,dt\,d^3\delta\vec{r}.
\label{eq:vx}
\end{equation}
This expression is very similar to the luminosity
formula~(\ref{eq:lumi}) except it contains the convolution with the
experimental vertex resolution $V^\mathrm{lab}(\delta\vec{r})$ and
there is no integration over $d^3\vec{r}$.  In analogy with van der
Meer method, we may integrate this equation over $d^2\Delta\vec{r}$ to
drop out $\rho^{\mathrm{lab}}_1$, and after deconvolution with
$V^{\mathrm{lab}}(\delta\vec{r})$ obtain the profile of the second
beam $\rho^{\mathrm{lab}}_2$. A possible non-zero beam crossing angle
complicates the formulas, but as it is shown below, it is always
possible to reconstruct the profiles of both beams {\it in their
  transverse planes} without any simplifying assumptions on the beam
shapes.

It is convenient to express the particle density of the second beam in
its own coordinate system $(x_2,\,y,\,z_2)=\vec{r}_2$ (see
Fig.~\ref{fig:coord}) as
\begin{equation}
  \label{eq:rho}
  \rho_2(\vec{r}_2,\, t) = \rho^\mathrm{lab}_2(\vec{r},\,t) = 
  \rho_2(\vec{r}_2^\perp ,\, z_2-|\vec{v}_2|t,\,0).
\end{equation}
The transformation Eqs.~(\ref{eq:frames}) define the relation between
the densities $\rho^\mathrm{lab}_2$ and $\rho_2$ in the laboratory and
in the second beam frames, respectively.  In the last equation we
distinguished transverse $\vec{r}^\perp_2 = (x_2, y)$ and longitudinal
$z_2$ coordinates and took into account that $z_2\parallel \vec{v}_2$.
In addition, let's define the resolution viewed from the second beam
frame as
\begin{equation}
  \label{eq:v2}
  V_2(\delta\vec{r}_2) = V^{\mathrm{lab}}(\delta\vec{r}).
\end{equation}

For the first beam we define
$\vec{R}=\vec{r}-\Delta\vec{r}-\vec{v}_1t$, so that
\begin{eqnarray}
  \label{eq:1}
  \rho^{\mathrm{lab}}_1(\vec{r}-\delta\vec{r}-\Delta\vec{r},\,t) =
  \rho^{\mathrm{lab}}_1(\vec{R}-\delta\vec{r},\,0).
\end{eqnarray}
Since $\int\!
\rho^{\mathrm{lab}}_1(\vec{R}-\delta\vec{r},\,0)\,d^3\vec{R}=1$, to
drop out $\rho^{\mathrm{lab}}_1$ from Eq.~(\ref{eq:vx}) one needs a
three-dimensional integration over $\vec{R}$.  The volume element
$d^3\vec{R}$ is invariant under rotations.  The simplest is to write
it in the coordinate system with $x$ and $y$ axes lying in the
$\Delta\vec{r}$ displacement plane and with $z$ axis pointing along
the unit normal $\vec{n}$. In components, $d^3\vec{R} =
dX^\Delta\,dY^\Delta\,dZ^\Delta$, where $\Delta$ superscript denotes
coordinates in this system.  Only $\rho^{\mathrm{lab}}_1$ in
Eq.~(\ref{eq:vx}) depends on $\Delta \vec{r}$, therefore regardless of
other variables, integration over $\Delta \vec{r}$ is equivalent to
integration of $\rho^{\mathrm{lab}}_1$ over
$dX^\Delta\,dY^\Delta$. The third coordinate $Z^\Delta = \vec{R}\cdot
\vec{n}$ is spanned when integrating over $t$ due to $|\vec{v}_1|t$
term in $\vec{R}$.

The density of the second beam, $\rho_2(\vec{r}_2^\perp ,\,
z_2-|\vec{v}_2|t,\,0)$ also depends on $t$, however. To decouple
$\rho_1^\mathrm{lab}$ and $\rho_2$, one may integrate in addition over
$z_2$ of the reconstructed vertexes. Then it is possible to change
the integration variables from $dt\,dz_2$ to $dZ^\Delta\,dZ_2$, where
$Z_2=z_2-|\vec{v}_2|t$.  $dZ^\Delta$ completes the integration over
$d^3\vec{R}$, while the integration over $dZ_2$ gives a projection
{\it transverse to the second beam}
\begin{eqnarray}
  \label{eq:proj2}
  \rho_2^\perp(\vec{r}^\perp_2) =\int\!\rho_2(\vec{r}^\perp_2,\,Z_2-\delta z_2,\,0)\,dZ_2=
  \int\!\rho_2(\vec{r}^\perp_2,\,z,\,0)\,dz.
\end{eqnarray}
Since $\vec{r}$ in $\vec{R}$ also depends on $z_2$ and
$\partial\vec{r}/\partial z_2 = \vec{v}_2/|\vec{v}_2|$, the Jacobian
of the variables substitution is
\begin{eqnarray}
  \label{eq:j}
  \left|\frac{\partial(t,\,z_2)}{\partial(Z^\Delta,\,Z_2)}\right| =
  \frac{1}{|\vec{n}\cdot(\vec{v}_1-\vec{v}_2)|} = \frac{1}{\cos\theta_z\cdot|\Delta\vec{v}|},  
\end{eqnarray}
where $\cos\theta_z$ is the same angle as in Eq.~(\ref{eq:vdm})
between $\vec{n}$ and $z\parallel (\vec{v}_1-\vec{v}_2)$.

Combining all pieces~(\ref{eq:vx})-(\ref{eq:j}) and (\ref{eq:gamma}) together, one has
\[\gamma_\perp\cos\theta_z\int\!\frac{1}{f \Delta T N_1 N_2\sigma}\cdot\frac{d^3 N_{\mathrm{vx}}}{dx_2\,dy_2\,dz_2}\,dz_2\,d^2\Delta\vec{r}=
\int\!\rho^{\mathrm{lab}}_1(\vec{R}-\delta\vec{r},\,0)\,d^3\vec{R}\,\times\]
\[\times\int\!\rho_2(\vec{r}^\perp_2-\delta\vec{r}^\perp_2,\,Z_2-\delta z_2,\,0)\,dZ_2\,
\int\!V_2(\delta\vec{r}^\perp_2,\,\delta z_2)\,d\delta
z_2\,d^2\delta\vec{r}^\perp_2=\]
\begin{equation}
  \label{eq:prof2}
  =\int\!\rho_2^\perp(\vec{r}^\perp_2-\delta\vec{r}^\perp_2)V_2^\perp(\delta\vec{r}^\perp_2)\,d^2\delta\vec{r}^\perp_2.
\end{equation}
Here were used the rotation invariance of the volume element
$dx\,dy\,dz=dx_2\,dy_2\,dz_2$ and the same for $d^3\delta \vec{r}$,
and also defined
\begin{equation}
  V_2^\perp(\delta\vec{r}^\perp_2) =
  \int\!V_2(\delta\vec{r}^\perp_2,\,\delta z_2)\,d\delta
  z_2=\int\!V^{\mathrm{lab}}(\delta\vec{r})\,d\delta z_2,
  \label{eq:vres2}
\end{equation}
which is the vertex resolution in the beam transverse plane.
Integration of Eq.~(\ref{eq:prof2}) over $\vec{r}^\perp_2$ gives the
total number of interactions and their rate consistent with
Eq.~(\ref{eq:int_any}).  Unfolding Eq.~(\ref{eq:prof2}) with
$V_2^\perp(\delta\vec{r}^\perp_2)$ gives the image of the second beam
in its transverse plane $\rho_2^\perp(\vec{r}^\perp_2)$.

From the very beginning it was assumed that only the first beam is
moved in van der Meer scan.  Instead, one can go to its ``rest'' frame
where vice versa, only the second beam moves by $-\Delta\vec{r}$ and
the first is stable. By repeating the procedure above, one can then
measure the transverse image of the {\it first} beam, as in this case
the integration over $\Delta\vec{r}$ and $z_1$ leads to the same
Eq.~(\ref{eq:prof2}) with the index 2 substituted by 1.  It is
interesting that two beam images can be obtained simultaneously from
the same set of vertex distributions measured at various
$\Delta\vec{r}$ points. They should be brought to the rest frame of
the corresponding beam, i.e. aligned differently during summation, and
projected to the beam transverse plane.

If $f$, $\Delta T$, $N_{1,2}$ or the reconstruction efficiency
entering $\sigma$ change during the scan, they should be considered as
functions of $\Delta\vec{r}$ during the integration on the left side
of Eq.~(\ref{eq:prof2}). On practice, this means that the accumulated
vertex distributions and their statistical errors should be reweighted
according to the factor $(f \Delta T N_1 N_2\sigma)^{-1}$ at every
scan point.

One can determine not only transverse but the full three-dimensional
images $\rho_{1,2}$, if the detector is able to measure the time $t$
of the interaction. Indeed, in this case Eq.~(\ref{eq:vx}) may be
integrated over $\Delta\vec{r}$ and $(\vec{r}-\vec{v}_1t)\cdot \vec{n}
= Z^\Delta$, while the integration over $Z_2=z_2-|\vec{v}_2|t$ may be
omitted.  This is possible since $Z^\Delta$ is directly measured. In
addition to the vertex resolution $V^\mathrm{lab}(\delta\vec{r})$, we
should include in the formulas the time resolution $V_t(\delta t)$,
make a change $t\to t-\delta t$ and integrate over $\delta t$. As
before, after the integration over $d^2\Delta\vec{r}\,dZ^\Delta$ or
$d^3\vec{R}$ the density $\rho_1^\mathrm{lab}$ drops out and we obtain
\[\gamma_\perp\cos\theta_z\int\!\frac{1}{f \Delta T N_1
  N_2\sigma}\cdot\frac{d^4
  N_{\mathrm{vx}}}{dx\,dy\,dz\,dt}\,d^2\Delta\vec{r}\,d([\vec{r}-\vec{v}_1t]\cdot
\vec{n})=\]
\begin{equation}
  \label{eq:threed}
  =\int\!\rho_2^\mathrm{lab}(\vec{r}-\delta\vec{r}-\vec{v}_2(t -\delta t),\,0)\,V^\mathrm{lab}(\delta\vec{r})\,V_t(\delta t)\,d^3\delta\vec{r}d\delta t.
\end{equation}
The distribution of vertexes in ($x_2$, $y_2$, $z_2-|\vec{v}_2|t$)
space, accumulated during the scan and deconvolved with the
resolutions determine the density $\rho_2$. Note, that the time
resolution affects the imaging in the longitudinal $z_2\parallel
\vec{v}_2$ direction. With an infinitely poor resolution in
$z_2-|\vec{v_2}|t$ measurement, Eq.~(\ref{eq:threed}) becomes
effectively equivalent to Eq.~(\ref{eq:prof2}).

Now let's consider the case when the particle distributions in $y$
direction and in $x$-$z$ crossing plane are independent.  In the
following, $x$,$z$- and $y$-dependent parts of a function will be
denoted by the corresponding subscripts, so that we have
$\rho^{\mathrm{lab}}_{1,2}(x,\,y,\,z,\,0) = \rho^{\mathrm{lab}}_{1,2\
  xz}(x,\,z) \cdot \rho^{\mathrm{lab}}_{1,2\ y}(y)$. As it was
discussed in Sec.~\ref{sec:lumi}, instead of scanning the full
$\Delta\vec{r}$ plane in this case it was sufficient to make two
one-dimensional scans over $\Delta x$ and $\Delta y$.  Instead of
$\Delta x$, any other line in $x$-$z$ plane may be taken, so for
generality we consider the line inclined at an angle $\theta_z$ from
the $x$ axis and denote the corresponding displacement as
$\Delta\vec{r}_{xz}= (\Delta r_{xz}\cdot\cos\theta_z,\,\,0,\,\,\Delta
r_{xz}\cdot\sin\theta_z)$.

If the spatial resolution also factorizes,
$V^\mathrm{lab}(\delta\vec{r}) = V^\mathrm{lab}_{xz}(\delta x,\,\delta
z) \cdot V^\mathrm{lab}_y(\delta y)$, according to Eq.~(\ref{eq:vx})
the vertex density can be split into two parts:
\begin{equation}
  \label{eq:rhosplit1}
  \frac{\gamma_\perp\cos\theta_z}{f \Delta T N_1 N_2\sigma}
  \frac{d^3 N_{\mathrm{vx}}}{dx\,dy\,dz}= 
  \rho^\mathrm{VX}(\vec{r},\,\Delta\vec{r})=
  \rho^\mathrm{VX}_{xz}(x,\,z,\,\Delta r_{xz})\cdot
  \rho^\mathrm{VX}_y(y,\,\Delta y).
\end{equation}
$\rho^\mathrm{VX}$ is the overlap integral from Eq.~(\ref{eq:vx}), its
normalization follows from Eq.~(\ref{eq:int_any})
\begin{equation}
  \label{eq:rhosplit2}
\int\!\rho^\mathrm{VX}(\vec{r},\,\Delta\vec{r})\,d^3\vec{r}\,d^2\Delta\vec{r}=1.
\end{equation}
Similarly, the relative normalization of $\rho^\mathrm{VX}_{xz}$ and
$\rho^\mathrm{VX}_y$ may be fixed by the requirement
\begin{equation}
  \int\!\rho^\mathrm{VX}_{xz}(x,\,z,\,\Delta r_{xz})\,dx\,dz\,d\Delta r_{xz} = \int\!\rho^\mathrm{VX}_y(y,\,\Delta y)\,dy\,d\Delta y = 1.
  \label{eq:rhosplit3}
\end{equation}
Following the same arguments as above one may then obtain the analogs
of Eq.~(\ref{eq:prof2}):
\[\int\!\rho^\mathrm{VX}_{xz}(x,\,z,\,\Delta r_{xz})\,dz_2\,d\Delta
r_{xz} = \int\!\rho_{2\ x2}(x_2-\delta x_2)V_{x2}(\delta x_2)\,d\delta
x_2,\]
\begin{equation}
  \int\!\rho^\mathrm{VX}_y(y,\,\Delta y)\,d\Delta y = 
  \int\!\rho_{2\ y}^\mathrm{lab}(y-\delta y)V_y^\mathrm{lab}(\delta y)\,d\delta y,
  \label{eq:rhosplit4}
\end{equation}
where $\rho_{2\ x2}$ ($\rho_{2\ y}^\mathrm{lab}$), $V_{x2}$
($V_y^\mathrm{lab}$) are the second beam transverse profile along
$x_2$ ($y$) and the corresponding projection of the resolution,
\begin{equation}
  \label{eq:rhosplit5}
  \rho_{2\ x2}(x_2)=\int\!\rho_{2\ xz}^\mathrm{lab}(x,\,z)\,dz_2,\ \ 
  V_{x2}(\delta x_2)=\int\!V_{xz}^\mathrm{lab}(\delta x,\,\delta z)\,d\delta z_2\ .
\end{equation}

If one integrates Eq.~(\ref{eq:rhosplit1}) over $d^3\vec{r}$, all
information about the spatial distribution of vertexes is lost and the
result should be expressible via rates.  To be consistent with the
normalization conditions~(\ref{eq:rhosplit3}) one should choose
\[  \int\!\rho^\mathrm{VX}_{xz}(x,\,z,\,\Delta r_{xz}^0)\,dx\,dz = 
\frac{R(\Delta r_{xz}^0,\, \Delta y_0)}{\int\!R(\Delta r_{xz},\,
  \Delta y_0)\,d \Delta r_{xz}},\]
\begin{equation}
  \label{eq:rhosplit6}
  \int\!\rho^\mathrm{VX}_y(y,\,\Delta y_0)\,dy = 
  \frac{R(\Delta r_{xz}^0,\,\Delta y_0)}{\int\!R(\Delta r_{xz}^0,\, \Delta y)\,d \Delta y},
\end{equation}
where $\Delta r_{xz}^0$ and $\Delta y_0$ are the fixed displacements
during the scan performed along the other coordinate.  As it was
pointed out in Sec.~\ref{sec:lumi}, they may be arbitrary.  Indeed,
since the rate is factorisable, $\Delta r_{xz}^0$ or $\Delta y_0$
appearing both in enumerator and in denominator cancel out.  Note,
that both Eqs.~(\ref{eq:rhosplit4}) and (\ref{eq:rhosplit6}) lead to
(\ref{eq:rhosplit3}) after integration over $x_2$, $y$, or over
$\Delta r_{xz}^0$, $\Delta y_0$, respectively.

Taking into account Eqs.~(\ref{eq:rhosplit4}) and
(\ref{eq:rhosplit6}), one can integrate (\ref{eq:rhosplit1}) over
$dy\,dz_2\,d\Delta r_{xz}$ or over $dx\,dz\,d\Delta y$ and finally
obtain
\[\int\!\frac{dN_\mathrm{VX}(\Delta r_{xz},\Delta y_0)}{dx_2}\frac{d\Delta r_{xz}}{N_0}
=\frac{R(\Delta r_{xz}^0,\Delta y_0)}{\int\!R(\Delta r_{xz}^0, \Delta
  y)\,d \Delta y} \int\!\rho_{2\ x2}(x_2-\delta x_2)V_{x2}(\delta
x_2)\,d\delta x_2,\]

\begin{equation}
  \label{eq:profsep}
  \int\!\frac{dN_\mathrm{VX}(\Delta r_{xz}^0,\Delta y)}{dy}\frac{d\Delta y}{N_0} =
  \frac{R(\Delta r_{xz}^0,\Delta y_0)}{\int\!R(\Delta r_{xz}, \Delta y_0)\,d \Delta r_{xz}}
  \int\!\rho_{2\ y}^\mathrm{lab}(y-\delta y)V_y^\mathrm{lab}(\delta y)\,d\delta y,
\end{equation}
where $N_0=\frac{f\Delta T N_1 N_2 \sigma}{\gamma_\perp \cos\theta_z}
= \int\!N_\mathrm{VX}(\Delta r_{xz},\,\Delta y)\,d\Delta
r_{xz}\,d\Delta y$.  One can see that the distribution of vertexes
accumulated during $\Delta r_{xz}$ ($\Delta y$) scan, projected to
$x_2$ ($y$) axis, normalised and unfolded with the corresponding
projection of the resolution $V_{x2}$ ($V_y^\mathrm{lab}$), gives the
transverse beam profile $\rho_{2\ x2}(x_2)$ ($\rho_{2\
  y}^\mathrm{lab}(y)$).

If the scan is performed not continuously but stepwise, so that the
data is taken at discrete points $\Delta x = \pm n\, \epsilon_x$,
$\Delta y = \pm m\, \epsilon_y$ with some integers $n$ and $m$, the
integrations may be approximated as discrete sums. This brings some
systematic uncertainty, which may be estimated in the end from the
reconstructed beam images.

\section{Discussion and conclusions}

The collider luminosity and cross sections can be measured in van der
Meer scan by sweeping the beams transversely across each other and by
measuring the collision rate as a function of the beam displacement.
We proved that the method remains applicable for the arbitrary beam
crossing angle and the beam displacement plane, see
Eqs.~(\ref{eq:int_any}),~(\ref{eq:res}) and (\ref{eq:genres}).

The four-dimensional beam densities may also be arbitrary.  In
particular, the formulas are valid for the beams initially mismatched
in time. With non-zero crossing angle the time shift between the beams
is equivalent to some shift in $\Delta x$, and therefore the
integration over $\Delta x$ in van der Meer method removes the
dependence on timing.  Also note, that in the case of $x$-$y$
factorization, when it is sufficient to scan only along two
perpendicular axes, their crossing point ($\Delta x_0$, $\Delta y_0$)
in Eq.~(\ref{eq:res}) is also arbitrary and not necessarily the point
of maximal luminosity. The beams at this point can be mismatched in
$\Delta x$ or, equivalently, in time.

The results are applicable to van der Meer scans performed at LHC
accelerator in 2010. This was a primary tool for the absolute
luminosity measurement at four major LHC experiments. The maximal
crossing angle of 540~$\mu$rad was in LHCb. It reduced the luminosity
by about 4\%,
%1./sqrt(1.+(0.27e-3*27.3e3*2/48.)**2)=1-0.044 - April scan, 
% 27.6 = szIP, 48 = svdM, 0.27e-3=540e-6/2 - only LHCb dipole
but caused a negligible correction, $\gamma_\perp-1 = 4\cdot10^{-8}$,
in van der Meer formula (\ref{eq:res}).
%1+(tan(0.27e-3))**2/2

The original van der Meer method suggested in 1968 was based on
counting the interactions. An important information is also contained
in the spatial distribution of vertexes. With excellent modern
detectors, like in LHC experiments, they are precisely measurable.  We
propose a new simple method how they can be used to reconstruct the
beam images. The vertex distributions transverse to the beam and
visible from its center should be accumulated during the scan and
unfolded with the transverse vertex resolution. This should give the
beam image in its transverse plane, see Eqs.~(\ref{eq:prof2}) and
(\ref{eq:profsep}).  The approach is valid for arbitrary beam shapes.

From the reconstructed normalized beam profiles one can determine
their overlap and then the luminosity using Eq.~(\ref{eq:split}).
This gives an alternative way of the absolute luminosity measurement
during van der Meer scan.  Two methods are independent as the beam
imaging method uses only the normalized {\it shapes} of the
accumulated vertex distributions, while the traditional method uses
only the {\it integrals}, i.e. the total number of interactions.

There is one complication in determining the luminosity from the
images, however. According to Eq.~(\ref{eq:split}), the luminosity
depends on the overlap integral of the beam profiles in $x$-$y$
laboratory plane,
$\rho^{\mathrm{lab},\,\perp}_{1,2}(\vec{r}_{\perp})$. With the
proposed imaging method one can reconstruct the profiles
$\rho^\perp_{1,2}(\vec{r}_{1,2}^\perp)$ transverse to the beams, see
Eq.~(\ref{eq:prof2}). They coincide with
$\rho^{\mathrm{lab},\,\perp}_{1,2}(\vec{r}_{\perp})$ only for the
collinear beams, but in general lead to a different overlap integral.

To correct for this effect, some information is needed on the
distribution of particles along the beam.  For example, in LHC one can
do the following. As it was pointed out, the non-zero crossing angle
reduced the luminosity and the overlap integral in 2010 scans by 4\%
or less. Since the effect is small, for its estimation it may be
sufficient to approximate the bunch shapes along $x_{1,2}$ and
$z_{1,2}$ as independent Gaussians with some effective sigmas
$\sigma_{x1,2}$ and $\sigma_{z1,2}$.  In this case the $x$-projections
in the laboratory frame are also Gaussian with sigmas
$\sigma_{1,2}^{\mathrm{lab},x}
=\sqrt{(\sigma_{x1,2}\cos\alpha_{1,2})^2+(\sigma_{z1,2}\sin\alpha_{1,2})^2}$.
If the bunch lengths of two beams are similar,
$\sigma_{z1}\approx\sigma_{z2}$, they can be obtained from the
$z$-width of the luminous region
$\sigma_{z1,2}\approx\sqrt{2}\sigma^z_\mathrm{lum}$. Here we assumed
collinear beams, but corrections due to non-zero crossing angle are
negligible since in LHC $\cos\alpha_{1,2}\approx1$ and transverse
sizes of the bunches are about three orders of magnitude smaller than
their lengths. Since for Gaussian beams the luminosity is proportional
to $\sqrt{\sigma_{x1}^2+\sigma_{x2}^2}$, its reduction due to the
crossing angle is
$\sqrt{\sigma_{x1}^2+\sigma_{x2}^2}/\sqrt{(\sigma_{x1}\cos\alpha_1)^2+(\sigma_{x2}\cos\alpha_2)^2+2(\sigma^z_\mathrm{lum})^2(\sin^2\alpha_1+\sin^2\alpha_2)}$.

If the detector is able to measure both the spatial coordinates and
the time of the interactions, one can reconstruct not only transverse
but the full three-dimensional beam images, see
Eq.~(\ref{eq:threed}). The luminosity can be determined from them
without any extra corrections or assumptions.

The proposed imaging during van der Meer scan is very similar to the
beam-gas imaging~\cite{massi}. The idea of the latter is to take a
beam ``photo''using the interactions with the gas remaining in the
beam pipe, assuming it is distributed uniformly in the transverse
plane.  In both methods, after deconvolution with the vertex
resolution, one can measure the transverse beam profiles and then the
luminosity, taking into account the crossing angle correction. The
beam-gas imaging method was successfully applied for the first time to
measure the absolute luminosity in LHCb~\cite{bg}.  Its accuracy, as
in van der Meer LHC scans in 2010, was dominated by errors in bunch
intensity $N_{1,2}$ measurement. The beam-gas method does not require
moving of the beams and can be used during normal data taking. On the
other hand, the advantage of the beam imaging during van der Meer scan
is a much higher statistics of interactions and a localization of
vertexes around a nominal luminous region where the vertex resolution
is optimal. The methods have different systematics, and it is very
advantageous to use both during van der Meer scan.

\end{document}